**Materials for Quantum Technologies: a Roadmap for Spin and Topology**

N. Banerjee[1*], C. Bell[2*], C. Ciccarelli[3*], T. Hesjedal[4*], F. Johnson[3*], H. Kurebayashi[5,6,7*], T. A. Moore[8*], C. Moutafis[9*], H. L. Stern[10*], I. J. Vera-Marun[11*], J. Wade[12*], C. Barton[13], M. R. Connolly[1], N. J. Curson[5], K. Fallon[14], A. J. Fisher[5,15], D. A. Gangloff[3], W. Griggs[9], E. Linfield[8], C. H. Marrows[8], A. Rossi[13,16], F. Schindler[1], J. Smith[17], T. Thomson[9], and O. Kazakova[13,18*&]

[1] Blackett Laboratory, Department of Physics, Imperial College London, London, SW7 2AZ, UK

[2] School of Physics, University of Bristol, Bristol, BS8 1TL, UK

[3] Cavendish Laboratory, University of Cambridge, Cambridge, CB3 0HE, UK

[4] Department of Physics, Clarendon Laboratory, University of Oxford, Oxford, OX1 3PU, UK

[5] London Centre for Nanotechnology, University College London, London, WC1H 0AH, UK

[6] Department of Electronic and Electrical Engineering, University College London, London, WC1E 7JE, UK

[7] WPI-AIMR, Tohoku University, 2-1-1, Katahira, Sendai 980-8577, Japan

[8] School of Physics and Astronomy, University of Leeds, Leeds, LS2 9JT, UK

[9] Department of Computer Science, University of Manchester, Manchester, M13 9PL, UK

[10] Department of Chemistry and Department of Physics, Photon Science Institute, University of Manchester, Manchester, M13 9PL, UK

[11] Department of Physics and Astronomy, University of Manchester, Manchester, M13 9PL, UK

[12] Department of Materials, Imperial College London, SW7 2AZ, UK

[13] National Physical Laboratory, Teddington, TW11 0LW, UK

[14] School of Physics and Astronomy, University of Glasgow, Glasgow, G12 8QQ, UK

[15] Department of Physics and Astronomy, University College London, London, WC1E 6BT, UK

[16] Department of Physics, SUPA, University of Strathclyde, Glasgow, G4 0NG, UK

[17] Department of Materials, University of Oxford, Oxford, OX1 3PH, UK

[18] Department of Electronic Engineering, University of Manchester, Manchester, M13 9PL, UK

* the authors have equally contributed to the article

[&] corresponding author

## Abstract

In this Perspective article, we explore some of the promising spin and topology material platforms (*e.g.* spins in semi- and superconductors, skyrmionic, topological and 2D materials) being developed for such quantum components as qubits, superconducting memories, sensing, and metrological standards and discuss their figures of merit. Spin- and topology-related quantum phenomena have several advantages, including high coherence time, topological protection and stability, low error rate, relative ease of engineering and control, simple initiation and read-out. However, the relevant technologies are at different stages of research and development, and here we discuss their state-of-the-art, potential applications, challenges and solutions.

### *Introduction and Figures of Merit*

Quantum physics has enabled the development of technologically useful components (*e.g.* transistors, lasers, magnetic tunnel junctions, etc.) that have transformed our economy and society. The next generation of Quantum Technologies (QTs) will be based on the physics of superposition and entanglement - this will require the development of new materials capable of supporting these phenomena. In this Perspective, we focus on material realisations of spin and topology as quantum objects for exploitation in future QTs, underpinning new perspectives for computing, sensing, communication, and information storage. These perspectives range from using the spin degree of freedom at the single spin limit, where optical and electronic control of isolated spins can achieve coherent control and spin manipulation with high fidelity, to concepts based on the subtle but powerful relativistic spin-orbit coupling, which have enabled several exciting breakthroughs, including topological spin textures in both real and momentum space. Magnetic skyrmions are a notable example; their topological protection enables an enormous stability at the nanoscale, leading to exciting proposals to use them as information carriers. Robust spin-textures also appear in momentum space in topological insulators that can produce highly efficient spin-to-charge conversion. Combining the physics of spin-orbit coupling and novel spin textures with superconductivity opens further synergies that harness the quantum mechanical phases of materials and generate new order parameters. Electronic spin qubits realised in both narrow and wide band gap semiconductors and in novel 2D materials already provide one of the most promising platforms for nanoscale optical communications networks and sensing.

Multiple national initiatives have set goals to achieve highly challenging quantum application targets. For example, the UK's National Quantum Strategy[1] has set the goal of a quantum-enabled economy by 2035, with *e.g.* a specific mission to realise quantum computers capable of running one trillion operations. The USA CHIPS and Science Act[2] looks to create a Quantum Network Infrastructure and develop critical standards. Canada's National Quantum Strategy[3] seeks to accelerate the development of quantum computing software, hardware, and algorithms. The EU Chips Act supports research and innovation initiatives in quantum technologies, in particular quantum chips for computing and sensing, to step up investment in frontier technologies. The European Declaration on Quantum Technologies[4] aims to create a world-class quantum ecosystem across EU member states and accelerate transition from "lab" to "fab" (*e.g.* improve scalability). In this context, development of various QT platforms has made significant progress, but still falls short of these envisaged targets.

| Host material | Level of maturity | Operational Temperature | Number of coupled qubits | Coherence Time | Readout method and speed | Benefits and Drawbacks | Refs |
|---|---|---|---|---|---|---|---|
| Defects & III-V nanostructres | Relatively high | 1.7 K - 300 K | 7 | ms - mins | Optical, 1 - 100 ns | ✓ High coherence times<br>✗ Deterministic creation | [1],[2],[3],[4],[5],[6] |
| Si/Ge QDs and Si donors | Relatively high | 4 K | 6 | ms - s | Electrical, < 1 ns | ✓ Established CMOS expertise<br>✗ Multi-qubit connectivity | [7],[8],[9],[10],[11] |
| Spins in superconductors | Relatively high | ~10 mK | 2 | ns | Microwave, 10 - 100 ns | ✓ Highly engineerable<br>✗ Noise and losses | [12] |
| Magnetic Skyrmions | Medium | *~ a few K* | *Concept stage* | µs | NV microscopy, scattering, TMR | ✓ nm-scale topological feature<br>✗ Reliable readout | [13],[14] [15] |
| Emerging 2D materials | Low | *~50 mK* | *Concept stage** | Not yet demonstrated | Electrical, 1 µs | ✓ Topological protection, valley degree of freedom<br>✗ Material quality and integration | [16],[17] [18] |
| Topological materials | Low | *~20 mK* | *Concept stage* | ns | Electrical or interference | | [19],[20] [21],[22] |

***Box 1: Key FOM for the materials platforms considered in this Perspective.*** *Italicised entries indicate theoretical/predicted values. * indicates ongoing initial experimental efforts. The Level of maturity is relative to the systems described in this Roadmap.*

We are at an exciting crossroads where breakthroughs in material science can enable disruptive technologies based on spin and topology. Specific topics covered by this Perspective article include semiconductor spin qubits, superconducting spintronics, magnetic skyrmions, two-dimensional (2D) and topological materials. We briefly address state of the art, main challenges and proposed solutions for each type of material systems and focus on their applications in QTs. Box 1 compares the materials platforms considered in this Perspective via a series of key Figures of Merit (FOM).

The *level of maturity* provides a qualitative assessment of the technological readiness of a particular material platform. It evaluates the ability to reproducibly fabricate materials in large quantities, whilst maintaining quality and quantum properties, and provides an indication of engineering challenges that must be overcome to scale to commercial viability.

The *operational temperature* represents the highest operational temperature for a given system. For semiconductor spins, it refers to the temperature range over which coherent control of spin qubits has been demonstrated. This temperature defines the infrastructure requirements of the quantum device; 10s of mK demands a dilution fridge, whilst a few K “only” needs a helium cryostat.

Future quantum computers and quantum networks will require connectivity of large numbers of qubits. The *number of coupled qubits* presents the maximum number of entangled qubits that have been demonstrated for each platform. For semiconductor systems, this may refer to entangled electron spins, or a register of coupled electron and nuclear spins. Skyrmions can be exploited as qubits when their size is of the order of nanometers. For example, Xia *et al.* proposes a skyrmion qubit scheme which, according to the authors, is scalable to process a large number of skyrmion connectivity using local electric field and spin current[5]. It is noteworthy that topological qubits (*e.g.*, Majorana zero mode (MZM)-based qubits[6], see Figure 4) require clear experimental demonstration of such states to be considered for viable quantum applications.

*Qubit coherence time* ($T_2$) gives an order-of-magnitude approximation of the spin coherence times reported for each platform. For semiconductor spin systems, the range represents electronic or nuclear spin coherence times measured by Spin Echo dynamical decoupling (DD) where the longest timescales are typically realised with nuclear spins (order of minutes). The coherence timescale sets the available time for performing quantum operations (*e.g.*, qubits for computing or networking applications) or describes the sensitivity (*e.g.*, for quantum sensing). Extending coherence times involves minimising external disturbances such as temperature fluctuations, electromagnetic noise, and material imperfections.[7]

*Readout method and speed* present the mode of qubit state measurement. Qubit readout should ideally be fast relative to the *coherence time*.

### ***Roadmap***

Our vision for the spin and topology roadmap for QTs is summarised in Box 2. Current and future material-related challenges, technology milestones, and critical infrastructure are shown in approximate chronological order. The selected material platforms (*i.e.* spins in semi- and superconductors, skyrmionic, topological and 2D materials) are organised according to the sections in this article, which are discussed in detail below. Each of them is expected to lead to applications with the advantages brought by spin and topology: high coherence time, topological protection and stability, lower error rate, ease of engineering and control, simple initiation and read-out. In this Perspective, we explore current and forthcoming applications of the relevant material systems in quantum computing, sensing, memory and metrology. We note that many of the technical challenges are applicable to the whole vast range of quantum materials and include scalable manufacturing pathways accompanied by supreme material quality, addressing decoherence mechanisms, accurate and deterministic control, advanced *in situ* and *in operando* characterisation, quantum metrology, readout, integration, and connectivity. Beyond device-level challenges, there is an urgent need to develop the quantum ecosystem, including the need for new infrastructure, standards, and regulations, as well as a focus on skills development and sustainability. The common challenges require development of quantum infrastructure, summarised as a horizontal middle section of the roadmap (Box 2).

QTs promise to bring enormous benefits to societies and economies around the world. Developments in materials science and engineering will play a fundamental role in the implementation and commercial exploitation of those technologies. Many nations have published their own Quantum Strategies developed in collaboration with industry, academia, and government with governmental investment of more about $44.5 billion announced to date at the time of preparing this Perspective.[8,9] The successful delivery of these strategies and commercialisation of QTs will require new funding mechanisms, innovations in engineering, and the development of national institutes and international collaboration focussed on materials for QTs.

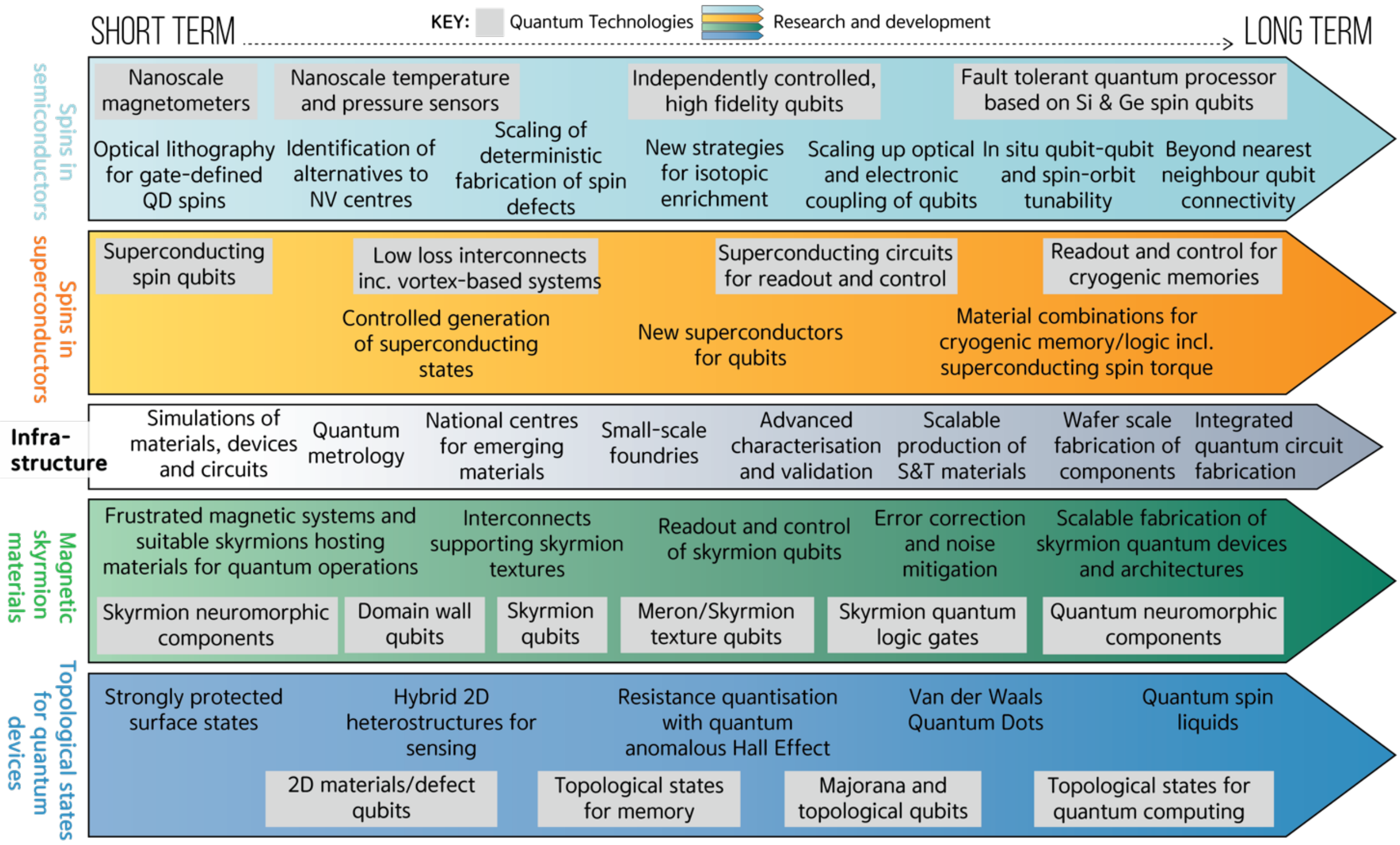

***Box 2: Materials for Quantum: Spin & Topology roadmap****. The coloured arrow shapes reflect the relevant R&D paths for each material type. The grey rectangles represent the targeted QT applications.*

### *Spin and topology-related phenomena in semiconductors*

#### State of the art

Semiconductors (narrow and wide bandgap) host a range of electronic spin qubits that can be incorporated in industrially mature solid-state devices with small physical footprints[10–13] (Box 1). Their main advantages for quantum technology include, *e.g.* incorporation of semiconductor spins in industrially mature solid-state devices with small physical footprints; optical connectivity enabling entanglement over long distances and isotopic purification giving access to long coherence times.

Control and readout of these qubits can be either electrical or optical, and the ability to control the qubit host platform chemically and isotopically[14] has led to the demonstration of spin coherence times over a minute.[15] Polarisation and control of hyperfine-coupled nuclear spins can give rise to multi-qubit quantum registers. Realisation of such registers has enabled the generation of entangled states of up to seven qubits with which the basic elements of fault-tolerant quantum logic have been demonstrated.[16] The existence of spin-dependent coherent optical transitions in some systems allows for long range distributed entanglement and arbitrary connectivity between qubit registers.

Spins in semiconductors span electrically controlled spins, including gate-defined silicon (Si) or germanium (Ge) quantum dots (QD) and single electron impurities in Si[12,13], and optically controlled spins, including self-assembled quantum dots in gallium arsenide (GaAs) or indium gallium arsenide (InGaAs) and defects in wide bandgap semiconductors[10,11]. In the latter, canonical systems include coherent spin defects in diamond, such as the nitrogen vacancy (NV) centre[17], as well as defects in silicon carbide (SiC)[18] and emerging materials, such as hexagonal boron nitride (hBN)[19,20] and gallium nitride (GaN)[21] (Figure 1a).

The overarching challenge facing these systems is the pursuit of deterministic spin qubit fabrication in a low-noise environment, whilst retaining high fidelity qubit control, feedback, and connectivity (Figure 1b). Advances in materials growth and device fabrication will dictate whether semiconductor spins can become leading platforms for QTs.

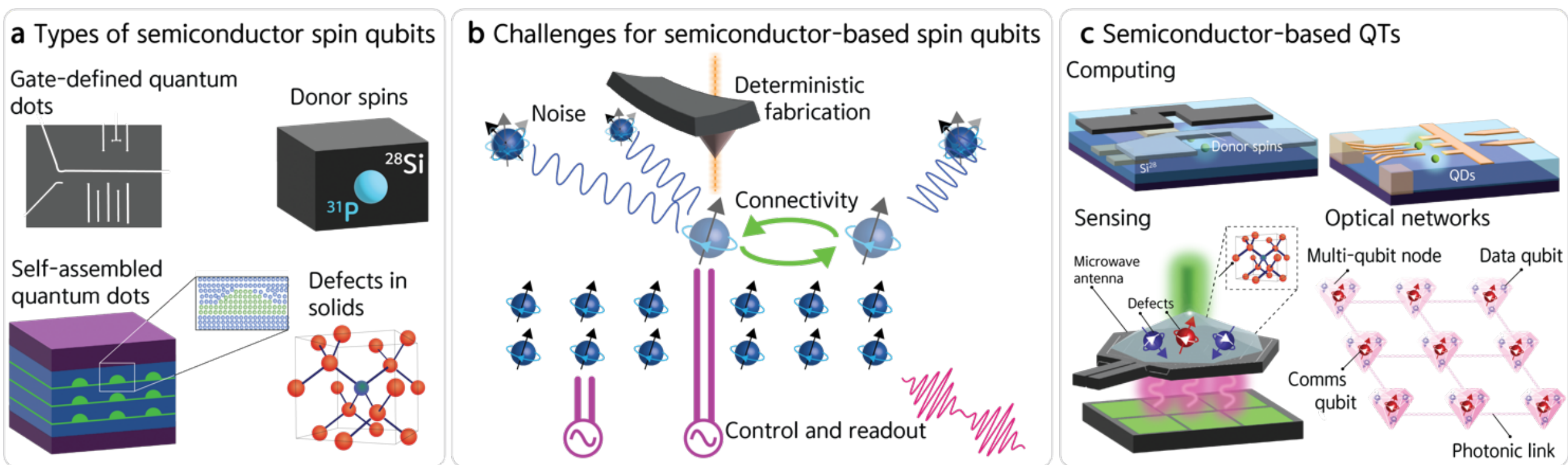

***Figure 1: Spin in semiconductors***. *(a) Schematic of the types of semiconductor spin qubits including gate-defined quantum dots, III-V self-assembled quantum dots, donor spins and defects in solids. (b) Schematic illustrating the challenges for spin qubits in semiconductors, including: their deterministic fabrication, minimising noise due to fluctuating magnetic and electric fields, optical and electrical control as well as readout and connectivity of multiple qubits. (c) Examples of quantum technologies enabled via semiconductor spins, including computation, sensing and quantum optical networks.*

### Applications in QT

Semiconductor spins represent a versatile variety of qubit platforms that offer applications across computation, simulation, sensing and communication[11,12,22]. The most commercially developed application of optically addressable solid-state spins is quantum sensing, including wide-field and spatially resolved quantum microscopy of biological and condensed matter systems. These applications have already been demonstrated with the nitrogen vacancy (NV) centre in diamond[23] and boron vacancy defects in hBN[24] (Figure 1c). In these examples, the spin acts as a sensitive probe of physical parameters, such as magnetic/electric fields, temperature or pressure, that can be 'read-out' optically.

Qubits with single- and two-qubit gate fidelities that are compatible with fault tolerant quantum computation (see Box 1 and Table S1 in Supplementary Information) have been demonstrated using gate-defined Si QDs[25,26]. Qubit arrays can be used to implement quantum simulation[27]. Nuclear spin states offer the extended quantum memory needed for error correction and fault-tolerant logic. Alternatively, optically addressable spin qubits such as in NV diamond and SiC offer quantum memory storage for future optical quantum networks, where stationary qubits (spins) couple to flying qubits (photons) as a resource for quantum communications and computing[28]. Efficient and coherent spin-photon interfaces, such as III-V QDs, offer the prospect of photonic entanglement in graph resource states for fusion-based quantum computing[29,30].

**Challenges and solutions**

*Fabrication*: Ideal solid-state qubits would be small, identical, and deterministically positioned in a defect-free material, making them easy to scale to large quantities. In practice, the degree of deterministic creation that has been achieved varies across qubit types. For single dopants in Si, ion implantation of spin-bearing species is a scalable and deterministic technology. For example, in-chip single ion detection and step-and-repeat high resolution implantation through the aperture of an Atomic Force Microscopy cantilever has been used to position phosphorus (P) dopants in Si (Figure 1b)[31]. Scanning tunnelling microscopy hydrogen resist lithography places dopants in Si with near atomic precision, and near-100% yield[31], although slow patterning speeds and CMOS compatibility remain significant challenges. For the scaling of gate-defined Si QDs, electron beam lithography fabrication is likely to be superseded by advanced optical lithography in industrial settings, demonstrating full compatibility with CMOS processes (see Roadmap, Box 2).

Deterministic creation of spin defects in wide bandgap materials has proven more challenging. Ion and electron beam irradiation can embed spin defects at depth in diamond with high yield, but laser-induced activation provides highest spatial precision and control[33]. For optically coupled systems, the requirement for spatial precision is likely to be lower than for electrically gated spins. A future requirement is to develop methods for high-yield generation of shallow coherent defects without compromising material quality and expanding these irradiation methods to encompass single defect creation in new systems (see Box 2).

*Noise minimisation*: Fluctuating magnetic and electric fields limit the accessible electronic coherence in semiconductors (Figure 1b). Common approaches to overcome nuclear spin noise include removal of nuclear spins altogether via isotopic purification[14], filtering the noise produced by nuclear spins via dynamic decoupling (DD)[34] or engineering of decoherence-free subspaces that are less sensitive to noise[35].

DD is used routinely for optically controlled III-V QDs and defects in wide-bandgap materials and has enabled the electronic spin coherence for single NV centres to be extended to seconds[36]. DD has been successfully applied to single spins in spin-rich materials (*e.g.* hBN) to achieve μs spin coherence times at room temperature[20]. Future work will demonstrate the extent to which nuclear spins can be decoupled in other such spin-rich systems.

Both Si and diamond are host materials with naturally low nuclear spin abundance, and growth of isotopically purified material is reasonably well-established, making them near-ideal hosts of spin qubits. Tuning the composition of nuclear isotopes has potential benefits in other materials, even if removal of nuclear spin is not possible (*e.g*. hBN and GaN) but can be expensive and highly research consuming. New methods of isotopic engineering (*e.g.* single ion implantation) are emerging in Si[37] and should be expanded and applied to new materials.

The degree of magnetic noise is known to be dependent on strain distributions, which can be controlled by precise material engineering. For III-V QDs, choice of low strain systems (GaAs/AlGaAs) has led to a 100x reduction in nuclear inhomogeneities, which was the main limiting factor for spin coherence[38]. Going beyond this will involve further reduction of material interface inhomogeneities, strain engineering, and active quantum control techniques for reducing nuclear spin noise amplitude.

Charge noise due to interface traps, crystal defects and dangling bonds (among other sources) is present in all solid-state systems. In Group IV hole qubit QDs, it has been proposed that hole-spin dephasing rates can be drastically reduced at sweet spots due to spin-orbit effects controlled by electric fields[39]. For defects in wide-bandgap materials, application of an electric field across defects can be employed for stabilisation.

*Control and readout*: Spin control and readout are based on spin resonance techniques: magnetic dipole or spin-orbit mediated, *i.e*. electrical or optical (Figure 1b). Scaling to large numbers of qubits requires a commensurately large number of control and readout lines – a considerable challenge, especially in cryogenic environments. Solid-state qubits typically suffer from inhomogeneity, which requires either broadband or tuneable control and readout technologies that add complexity. In multi-qubit systems where conditional logic is based on proximal interactions, high fidelity control and readout often clash with high-fidelity multi-qubit gates due to crosstalk and nanofabrication restrictions in high-density systems. In gate-defined or donor systems with multiple proximally coupled spins, these requirements lead to large device footprints. In systems where multiple qubit types (optical, electronic, and nuclear) are employed, there is an additional challenge of hybridisation, of, *e.g.*, integrated optics with microwave antennas, and of operating and synchronising coherent sources across a wide range of the electromagnetic spectrum[40]. In many qubit systems, spin-orbit coupling is required to perform qubit control[41], but introduces relaxation and dephasing via, *e.g.*, phonon coupling.

Progress on the above challenges will come from a focused effort at standardisation of classical interface technologies (*e.g.*, micro-optical or radiofrequency resonators, multiplexed sources, and switches) that can operate on the timescales set by qubit coherence. High-efficiency single photon detectors and low-noise high-bandwidth cryogenic amplifiers will be critical in achieving high-fidelity readout on timescales allowing feedback and error-correction. On the qubit side, the ability to tune spin-orbit and qubit-qubit interactions *in-situ* could also be a key to achieving high-fidelity control together with long coherence time (see Box 2).

*Connectivity*: Application of semiconductor spins to quantum computation, communication and simulation will require connecting multiple qubits, which will be either nearest neighbour or via photons (Figure 1b). For nearest neighbour connectivity, the difficulty of this challenge depends on the host material (in particular, its band gap) and the position of the bound states within the bandgap. For shallow donors (or acceptors) and gate defined QDs, the binding energies can be modulated by additional gates that raise or lower the effective tunnel barrier. The exchange interaction between carriers then decays over comparable distances and can be used to execute two-qubit gates between adjacent qubits using native spin encoding. The current state of the art is a linear array of 6 Si/SiGe QDs and a two-dimensional array of 4 Ge QDs[42] (see Table 1 in Supplementary Information). While it is important to go beyond nearest-neighbour connectivity, this is very challenging to achieve, since the exponential decay of the exchange is determined by that of the wavefunction. One possibility is physical shuttling of the electron to a remote location[43]. Another is to couple the spins on-chip via a microwave cavity, utilising an intrinsic or induced spin-orbit interaction[44].

Colour centres will require optical connectivity as the wavefunctions are highly localised. Therefore, longer-distance interactions rely on the same optical interface that is used for qubit control and remote entanglement of NV centres, with sufficiently strong zero-phonon lines to enable useful entanglement rates via optical cavities. These have been demonstrated over km-scale distances using probabilistic quantum erasure[45]. Poor emitter quantum efficiencies, outcoupling, and telecom conversion currently limit the entanglement rates to ~Hz. An alternative approach to connectivity involves a 'broker qubit' that interacts with a register, *e.g.*, colour centres and QDs can be coupled to a local register of many nuclear spins using hyperfine contact or dipolar interactions[46].

**Perspective**

Spins in semiconductors are well placed to offer a source of spin qubits in a scalable platform. Their solid-state nature is both a benefit and a limitation – it acts as a source of spin and charge noise and provides significant challenges for deterministic spin qubit creation. The future prospects of semiconductor spin qubits will rely on the coordinated collaboration across quantum physics, electrical engineering and material science. The effort across these areas of expertise should be focussed on both enhancing existing systems and identification and development of spin systems in novel materials.

***Spin and topology-related phenomena in superconductors***

**State of the art**

A key use of superconductivity in quantum devices is the possibility of creating macroscopic numbers of phase coherent electrons – in the form of Cooper pairs – which are generally robust to non-magnetic disorder. Thin films of conventional superconductors can therefore be patterned into devices without degrading the superconducting transition temperature. Superconductors can be used to build non-linear devices such as Josephson junctions, inductors and capacitors, as well as low resistance interconnects. Using these elements one can create resonant circuits, most notably giving rise to the concept of 'superconducting artificial atoms' which can form the basis of qubits, for example[47]. As a consequence, the use of conventional s-wave singlet superconductivity in QTs is well established across diverse areas, from superconducting qubits[48] to nanowires for single photon detection[49]. Generating new functionalities, with emerging materials and combinations that utilise spin and topology, represents novel disruptive directions[50]. The current state of the art is spread across a range of systems, including superconductor-semiconductor architectures and fully metallic systems.

Odd-frequency, spin-triplet ($S$ = 1) superconductivity[51] is an example of emergent physics in thin film heterostructures. This exotic state can be generated with various types of spin mixing, including spin-orbit coupling[52]. The emerging functionalities include such examples as vortex-based spin-interconnects[53], π-shifters[55], superconducting diodes[56,57], Andreev/Majorana qubits[58], supercurrent torques induced by spin-triplet supercurrents[59], and reprogrammable magnon-fluxon interaction[60] (Figure 2).

Key advantages of these novel Cooper pairs are robustness against pair breaking via the exchange field in neighbouring magnetic layers and ability to carry spin in contrast to conventional singlet pairs. Such engineered properties have the potential to impact high-level components in computing architectures: for example, superconducting spin-based processors and memories, and superconducting interconnects with vortices mediating spin information for relevant applications in QTs[53,54].

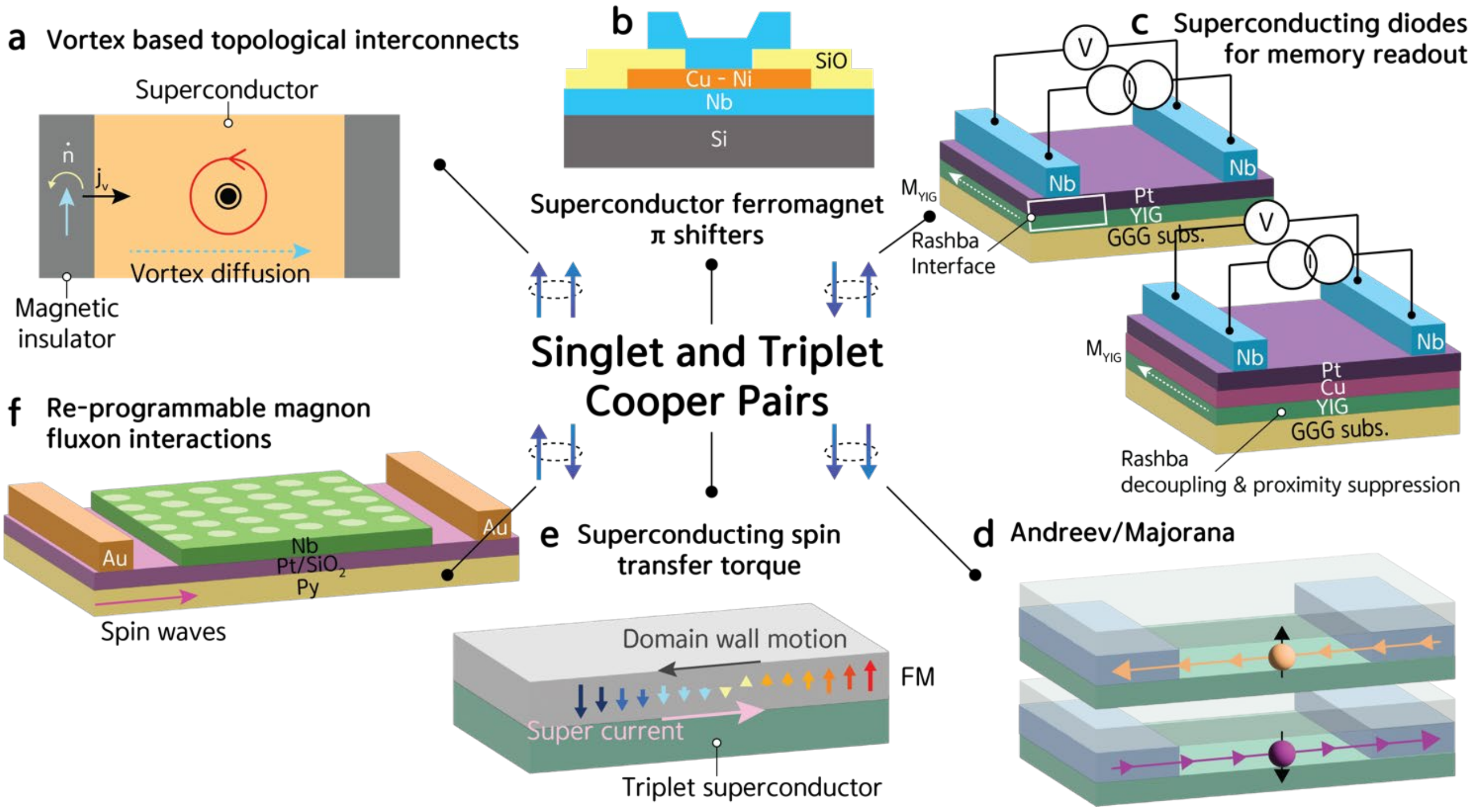

***Figure 2: Emergent functionalities in superconducting structures for topological and ferromagnet-based processing, interconnects, and memory****. (a) vortex-based spin-interconnects, (b) π-shifters, (c) superconducting diodes, (d) Andreev/Majorana qubits, (e) supercurrent torques induced by spin-triplet supercurrents, and (f) reprogrammable magnon-fluxon interaction.*

### Applications in QT

From the memory perspective, realistic scaling of superconducting logic and quantum computers requires, among other advances, non-volatile classical memories. These memories need to be energy efficient at low operational temperatures, primarily due to the low cooling powers available.[61-63] Magnetic memories remain among the most promising routes to robust memory for QTs[64,61], but at low temperature, the energy targets become more stringent because of the lower cooling powers. The upper limit for the writing energy in a cryogenic memory can be estimated from the achievable cooling powers to be in the range of 10 fJ/bit, considering a density of $10^8$ bit/cm$^2$ and a bit write/read time of 1 ns. This is comparable with what is currently achieved in metallic spin orbit torque MRAMs[65,66]. However, fundamentally lower energy consumption in magnetic memories at low temperature is possible by suppressing Joule heating. The interface with superconductors offers completely new energy-efficient venues for reading cryogenic magnetic memory[67], and potentially switching the bits via superconducting torque[59], although the latter remains to be demonstrated. The superconducting condensation energy stored in 100 nm$^3$ of Nb is ≈10 aJ[68], much higher than the 60 $k_BT$ energy barrier that a ferromagnet needs at 4.2 K to switch from one bit-state to the other. Finding new ways to use this condensation energy in the superconductor released during the transition to do work on nearby magnets is a key current challenge. Ultimately applications with the highest speeds will require extending these basic studies to other magnetic materials (*e.g.*, antiferromagnets).

On the processor side, superconducting qubits based on conventional s-wave superconductors typically exhibit lifetimes in the range of 10–100 μs, constrained by charge/flux noise and two-level system losses.[69] These lifetimes limit two-qubit gate fidelities and necessitate the use of hundreds of thousands of physical qubits to implement quantum error correction. Several benefits are expected from novel materials. The development of π-junction shifters, notably demonstrated using ferromagnetic weak and strong barriers[70-72] could reduce the footprint of conventional qubits and realise flux qubits without the need for external flux bias.[73] Realising a spinless p-wave superconductor is expected to host a degenerate ground state of quasiparticles with non-Abelian exchange statistics required for topologically protected quantum computing.[74] Although there is evidence for bulk p-wave superconductivity in a range of exotic materials[75], realising these bulk materials as thin films suitable for devices is challenging (*e.g.* Ref. 76). Utilising proximity effects are more flexible: superconductor-semiconductor hybrids are a leading platform for realising such p-wave superconductivity but require semiconductors with strong spin-orbit coupling and the application of high external magnetic fields. This drive to realise topological qubits has also enabled the development of novel qubits based on single quasiparticle spins[77] and artificial Kitaev chains[78].

**Challenges and solutions**

*Scalable material platforms*: The future development of superconducting-based QTs depends on our ability to controllably generate unconventional superconducting states with scalable materials platforms that can be integrated with existing technologies. Significant research has gone into designing superconductors and semiconductors and their hybrids with minimal disorder as disorder-induced sub-gap states are detrimental to induced topological superconductivity. While gate control of spin-orbit coupling is an obvious advantage, it remains a challenge to reliably deposit the gate dielectric. A second challenge in this area is identifying superconductors with high critical fields and temperatures. While elemental superconductors such as Al, Nb, and NbN provide a well-studied and stable foundation, the low critical temperature and field are limiting factors.

Completely disruptive systems, *e.g.* thin films of high temperature or unconventional superconductors ($MgB_2$, cuprates, topological superconductors), are more challenging to fabricate and integrate into devices. By contrast, existing systems (*e.g.* Al, Nb, NbN) can serve as a building block, providing a familiar foundation for growth and device fabrication, on top of which novel physics can be achieved with heterostructuring (various examples in Figure 2) to enhance their desirable properties.

*Decoherence mechanisms:* Beyond the problem of materials selection, understanding decoherence mechanisms in these devices also possess significant challenges. This also requires advancements of characterisation techniques, especially *in situ* techniques, which can probe materials under operating conditions of voltage and/or temperature bias. Physical constraints (*e.g.* electron transparency at heterointerfaces) also present fundamental limits to the degree of coupling between different types of electronic states, which is an issue in many device architectures combining disparate materials systems, motivating continuous exploration of such architectures and interfacial engineering.

Future progress will come from understanding material-related constraints with existing materials and parallel efforts to discover new material systems. Demonstrating functional devices utilising the spin degree of freedom in the superconducting state and understanding their operation in the dynamic state are already underway.[54] In this context, the field of superconductor/semiconductor devices with emergent topological phases is more mature but with its own challenges as discussed above. Addressing these challenges will underpin the realisation of scalable and fault-tolerant superconducting QTs.

**Perspectives**

Superconductivity is already established as a key component in various QTs. The materials currently in use are conventional, singlet, superconductors with no inherent topological properties (*e.g.* Al, Nb, NbN). The use of new materials or existing material combinations in the form of thin film hybrids to engineer unconventional emergent phases opens the door to many new spin- and topology-based avenues for technological exploitation[79-81]. Superconductor/ferromagnet hybrids, especially those with spin-orbit coupling, allow gate-tunable spin-based functionalities via the generation of triplet Cooper pairs, provide effective platforms for static and dynamic memory applications.[54] Through these advancements, we expect entirely new avenues of computing together with significant reductions in energy consumption and improvements in system efficiency by linking processors and memory.

### *Magnetic Skyrmions*

#### State of the art

Skyrmions are topologically non-trivial spin textures with particle-like properties[82]. Their small size, topological stability, and ease of manipulation with spin-torques make them ideal for spintronic-based applications[83]. Initially discovered in materials with chiral B20 lattices at low temperatures (*e.g.* MnSi[84], they are now widely explored in room-temperature metallic multilayers with interface-induced chiral interactions[85]. Figure 3 illustrates the current landscape of emerging skyrmionic QTs from materials development to device design, promising advances in gate-based quantum computing using qubits based on skyrmion helicity or meron polarity, as well as topological quantum computing when coupled to superconductors.

Whilst classical skyrmionics is well-established in terms of materials, devices, and potential applications, the subfield of quantum skyrmions is still at a very early stage, with key concepts formulated in theory but still to be subjected to experimental tests owing to the extreme challenges of nucleating, controlling and detecting skyrmions at suitably small (nm-scale) sizes.[86] Nevertheless, work to date shows the promise of skyrmions as a unique platform that supports novel quantum phenomena in nanoscale spin systems, challenging orthodox notions of topology, offering a fusion of spintronics, quantum information, and strongly correlated systems, and opening the prospect of performing quantum information processing and storage in the same device.

The main advantages of skyrmionics systems for quantum applications include i) topological stability, *i.e.* robustness against perturbations such as defects, thermal fluctuations, and external noise, making skyrmions desirable in quantum systems, where decoherence and susceptibility to local disturbance are major challenges; ii) nanoscale size which is useful for quantum computing and memory applications where miniaturisation is critical; and iii) low energy manipulation for reduced energy consumption and efficient control mechanisms.

Additionally, Box 1 and Table S2 (Supplementary Information) summarise the FOMs and the current state of the art for the key quantum parameters in skyrmionic systems.

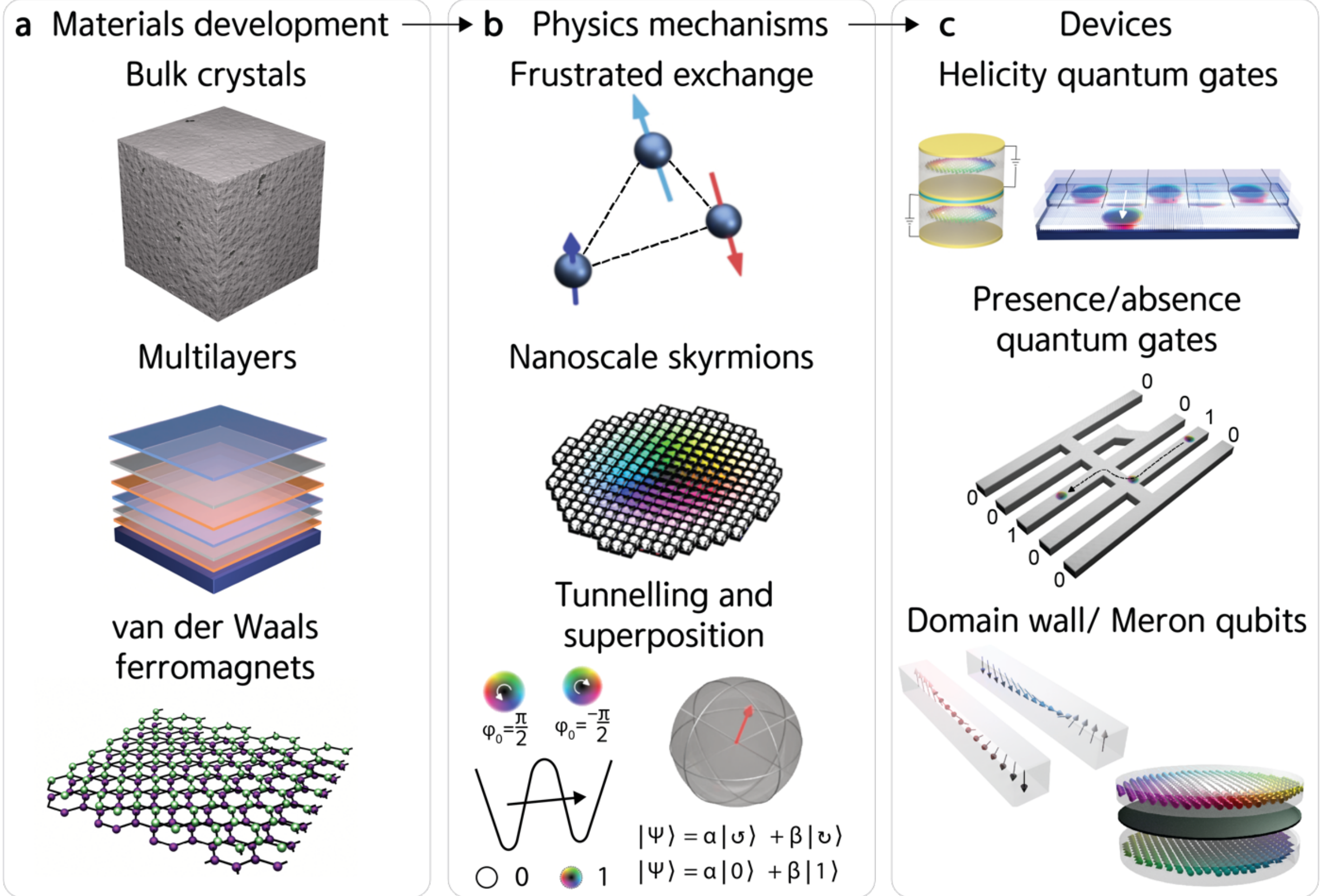

***Figure 3: From materials to different types of proposed skyrmion qubit devices.*** *(a) Materials development: bulk, multilayers and 2D systems supporting skyrmions. (b) Physical mechanisms: frustrated exchange interactions leading to stabilisation of skyrmions at the nanoscale and unlocking a new class of skyrmionic devices, which leverage quantum mechanical effects, including superposition and tunnelling. (c) Device design and verification/validation: quantum information encoding utilises nanoscale skyrmions' presence/absence (middle) or helicity. Designs of skyrmionic quantum gates based on both static and 'flying' qubits have been proposed (top). Similar developments in material engineering and device design incorporating domain walls or merons (half-skyrmions) for encoding quantum information (bottom).*

## Applications in QT

Skyrmionic spin textures have been proposed for skyrmion-based Boolean computing[87] and neuromorphic computing[88]. Many of their advantages in those contexts are equally applicable in quantum computing, setting skyrmions as quantum logic elements[5,87,89–92].

Their quantum properties (*e.g.* tunnelling and superposition) pave the way for applications in quantum computing, notably skyrmionic qubits and qubits based on merons ("half skyrmions")[90,91] (Figure 3c). Psaroudaki *et al.*[90] recently introduced skyrmion-based quantum logic elements, where qubit information is encoded in helicity and manipulated with electric and magnetic fields to perform gate-based quantum computing. Skyrmion-based qubits can be controlled by microwave fields and readout using non-volatile techniques, making it possible to scale to large quantities. Frustrated magnets can be used to stabilise the quantum skyrmion spin texture[93] yielding degenerate skyrmion helicities that can be superposed to form a qubit[90]. Tuning the external helicity potential enables the observation of a range of quantum effects, including macroscopic quantum tunnelling, macroscopic quantum coherence and macroscopic quantum oscillation. Recent research suggests using the core spin direction in nanoscale merons within magnetic nanodisk as qubits[91]. Core spin directions are assigned to be the qubit states $|0\rangle$ and $|1\rangle$, which enables the construction of the three quantum gates necessary for universal quantum computation (Figure 3b). Moreover, skyrmion/superconductor heterostructures offer the potential of topological quantum computing[94-96].

### Challenges and solutions

*Materials and Fabrication:* Key challenges involve the identification of optimal materials for skyrmion quantum applications that are compatible with industry preparation methods. Whilst conventional skyrmion multilayer materials are CMOS back-end-of-line compatible, this may not be the case for some of the novel materials required here. The search spans from single crystals and polycrystalline powders to 2D van der Waals (vdW) materials, including frustrated magnets with noncollinear spin textures that manifest the helicity degree of freedom. Bloch-type nanoscale skyrmions have been identified in $Gd_2PdSi_3$, a centrosymmetric triangular-lattice magnet, at low temperature[97,98]. Potential hosts for frustrated skyrmions include triangular magnets with transition metal ions (*e.g.*, $NiGa_2S_4$, α-$NaFeO_2$, $Fe_xNi_{1-x}Br_2$ dihalides) and the frustrated square lattice $Pb_2VO(PO_4)_2$ with its ferromagnetic nearest-neighbour interactions[5]. Recent theoretical advances encourage the development of bilayer systems using frustrated magnets (*e.g.* vdW ferromagnets) to host skyrmions, moving beyond single-layer or bulk investigations[5]. Advances in device nanofabrication are crucial, *e.g.*, in the identification of bilayer nanodiscs small enough to support neighbouring coupled skyrmionic states with minimum nanometer-wide size stability.

*Device-level Integration and Operation:* Concerted efforts are needed on core system-level challenges such as readout mechanisms, integration with existing QTs and classical peripherals, scalability, error correction and noise. The scientific community is just beginning to tackle these issues, whilst simultaneously exploring skyrmion QT applications[5,89–91,99]. There are also challenges particular to specific device concepts, *e.g.*, temperature control for the stable operation of skyrmionic quantum systems and precise control of skyrmion position. The latter is particularly important in certain contexts, *e.g.* the Ising coupling gate with the proposed use of a square grid pinning pattern, which is challenging to fabricate on the nanoscale (Figure 3)[5].

*Control and Readout:* It has been predicted that, while qubit control is typically achieved with external magnetic fields[90], control of qubit helicity can be realised with electric fields and spin currents.[5] Reliable readout is essential for skyrmion-based quantum-computing. NV microscopy can discern qubit helicity. Resonant elastic x-ray scattering, and ferromagnetic resonance techniques enable skyrmion helicity observations and single qubit readout[99], while magnetic force microscopy resonators detect magnetic states through frequency shifts[5,91]. Recent proposals include using tunnel magnetoresistance for helicity skyrmion qubit readouts at device level[5]. Achieving a dependable, non-volatile device-level readout remains a significant challenge. Figure 3 illustrates the current landscape of emerging skyrmionic QTs from materials development to device design.

Critical FOMs, which will need to be optimised for successful development of skyrmion-based quantum technologies, include qubit thermal stability, size, coherence time, gate operational time, and initialisation/readout energy (see Box 1). These FOMs are compared across the literature in Table S2 (Supplementary Information).

### Perspective

In advancing skyrmion-based QT, the synergy of understanding and fine-tuning of physical systems (*e.g.*, nanoscale skyrmions) becomes crucial, aligning with quantum hardware demands and tolerances. For example, in quantum neural networks, *e.g.*, quantum reservoir computing, decoherence may enhance system nonlinearity, offering greater noise resilience than traditional gate-based approaches.

## *Topological and Emerging 2D Materials*

### State of the art

Topological materials promise to channel the notion of topological stability, *i.e.*, invariance under continuous deformations, which can be harnessed for desirable electronic, spintronic, and quantum applications. The 'shape' in question does not usually refer to a geometrical shape (see, however, Section on Skyrmions), but instead to the quantum mechanical wave function of the material many-body ground state. The main advantages of topological systems for quantum applications include robustness to disorders and defects enabling fault-tolerant quantum computation, potential for room temperature operation, whereas dissipationless transport is beneficial for low-power quantum circuits.

When electronic interactions are weak, the electronic part of the ground state can be approximated by a product over single-electron wave functions labelled by crystal momentum. This unlocks a wealth of topological invariants, capitalising on the fact that crystal momentum is defined on a (d-dimensional) torus. The resulting topological band theory predicts topological insulators (TIs) hosting lossless edge states[100] (Figure 4a), as well as topological semimetals with relativistically dispersing quasiparticles and surface Fermi arcs[101]. A famous example of the use of a magnetic TI is the integer quantum Hall effect (Figure 4c) that has revolutionised metrology[102]. In the guise of the Bogoliubov-de-Gennes mean-field formalism, the theory furthermore predicts topological superconductors with exotic Majorana edge states[103] (Figure 4b), that could be used for error-free quantum computation[104].

Topological band structures support boundary modes protected by symmetries or Chern numbers. In 2D, these include chiral (QAH) or helical (QSH) edge modes; in superconductors, effective p-wave pairing can host MZMs[105]. In 2D van der Waals semiconductors and graphene, weak hyperfine coupling and tunable spin-orbit fields enable spin/valley qubits and engineered proximity effects. Below, we quantify where these properties translate into metrology-grade quantisation, low-error logic, or robust interconnects.

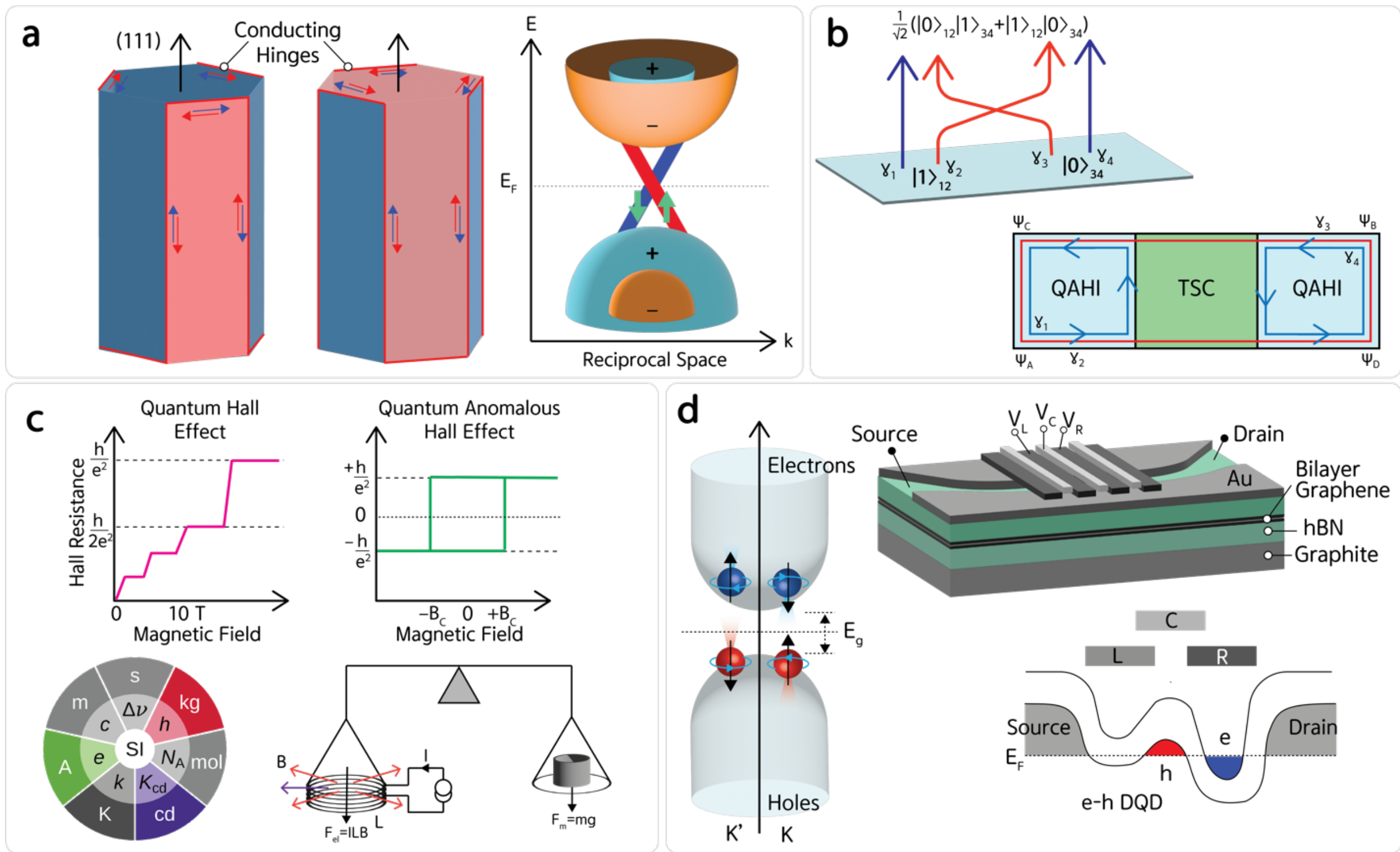

***Figure 4: Quantum phenomena in topological and emerging materials***. *(a) Illustration of the symmetry-allowed hinge state configurations of a higher-order topological insulator (HOTI) with inversion symmetry. (b) Proposal for braiding MZMs in a quantum anomalous Hall insulator (QAHI) - topological superconductor (TSC) - QAHI junction. (c) Schematic plots of the Hall resistance as a function of applied magnetic field for the quantum Hall effect (left) and the anomalous quantum Hall effect (right), showing fully quantized zero-field resistance for the latter. The QAHE is a contender for not only realising the ampere but can contribute to the realisation of other SI units (e.g. the kilogram, using a Kibble balance that does not*

*require large magnetic fields, and the candela, via cryogenic radiometry). (d) Band structure of bilayer graphene showing the presence of valley-dependent orbital magnetic moments (left) and schematic representation of a hBN-encapsulated bilayer graphene structure (right) showing the electrostatic formation of a quantum dot.*

## Applications in QTs

A promising application based on topological materials is high-temperature, low-magnetic-field resistance quantisation utilising the quantum anomalous Hall effect (QAHE)[106]. Due to an internal magnetic field arising from the spin-orbit interaction in a magnetic system, resistance quantisation in terms of $h/e^2$ has been experimentally demonstrated in the absence of an external magnetic field, which opens the door to quantum metrology applications. The Quantum Hall Effect (QHE, Figure 4c) requires a large magnetic field for such quantisation, whereas QAHE can display its quantisation as a remnant (zero field) property. $(Bi,Sb)_2Te_3$ with Cr (and V) doping is one of the most successful strategies in simultaneously achieving long-range magnetic order, low bulk carrier densities, and topological electronic transport[106-108]. QAHE resistance quantisation with an accuracy of one ppm near zero magnetic fields was recently demonstrated[109]. This was subsequently improved by further material optimisation to 0.01 ppm (at a modest field of 0.2 T).

Beyond QAHE, quantum spin Hall (QSH) insulators offer helical edge channels without net magnetisation. Proximitising QSH edges with an s-wave superconductor yields an effective spinless 1D channel and a route to MZMs without large external fields. As an alternative to Rashba nanowires, intrinsically topological materials interfaced with superconductors *(e.g.*, magnetic TI/QSH platforms) provide cleaner band topology and stronger spin-orbit coupling, potentially lifting topological gap energies and easing braiding and readout. Recently, QSH states have been detected at zero external magnetic field in graphene proximitised with a 2D antiferromagnet[110], opening a route for practical applications of 2D heterostructures in quantum spintronic circuitries.

2D materials offer quantum confinement via their inherent thinness and excellent electrostatics. They exhibit high-quality structures with pristine vdW interfaces, fulfilling some of the material requirements for realising solid-state quantum computing[111]. 2D vdW QDs hold great promise in solid-state quantum computation[112] (Figure 4d). In contrast to Si and GaAs systems discussed earlier, the low spin-orbit coupling and negligible hyperfine coupling result in a long spin-coherence time in graphene[113], making it ideal for spin memory, whereas transition metal dichalcogenides (TMDs) with large spin-orbit coupling offer fast spin-qubit operation. Assembling heterostructures from a variety of 2D materials is an effective approach for designing solid-state qubits with optimised device parameters. The recent discovery of superconductivity in twisted bilayer graphene has galvanised research into the interplay of electron correlation and band topology[108]. This offers the opportunity to tune band topology on demand to induce exotic phases of matter such as high-temperature superconductivity and fractional Chern insulators. Solid state qubits in 2D materials would efficiently couple with photonic qubits generated by single-photon emitters (*e.g.,* wide-bandgap TMDs and hBN, see Section 2), opening avenues for distributed quantum networks.

In MZM qubits, information is stored nonlocally in the joint fermion parity of spatially separated MZMs, whereas Clifford operations arise from braiding, with measurement-based schemes enabling universality. Readout leverages parity-to-charge conversion, microwave dispersive shifts, or teleportation-style protocols. Key metrics are: (i) an induced topological gap large in respect to $k_B T$ and charge noise; (ii) poisoning times exceeding total gate plus measurement time; (iii) unambiguous detection (beyond the topological gap protocol (TGP)) and, ultimately, braiding demonstrations. Current gaps in proximitized semiconductors are typically ≤ ~50 µeV, keeping the field in the exploratory regime.

In the long term, another promising platform for quantum computation is quantum spin liquids that rely on frustrated magnetism[114]. Similar to topological superconductors, quantum spin liquids can host fractionalised excitations (anyons), particularly MZMs which can be envisaged as a unique quantum state that connects the two ends of a 1D system or two vortices[115]. Making use of the non-local nature of these excitations, braiding MZMs perform topologically protected quantum operations that are robust against local perturbations (Figure 4b). Creating and controlling stable quantum spin liquids as well as topological superconductors in real materials remain a major challenge.

**Challenges and solutions**

*From lab demonstration to practical devices:* So far, QAHE measurements have been performed in cryogenic systems in research labs. For more practical use, becoming independent from bulky, expensive superconducting magnets that perform extremely slowly is a big step forward for QAHE in metrology applications to define the standard units of resistance and current, as well as the electrical realisation of the kilogram via the Kibble balance (Figure 4c). The effective use of the internal magnetic fields (*i.e.* magnetic anisotropies) in magnetic TIs can potentially solve this issue[98].

*Material quality*: The need of QAHE for mK temperatures requiring dilution refrigerators restricts its use for a certain range of quantum electronics applications. This temperature limit is not a fundamental barrier but determined by materials issues such as a lack of control over the dopants, growth process and defect formation, which lead to detrimental bulk conduction and loss of precise quantisation[109]. Indeed, QAHE at 1.4 K has been reported in the layered van der Waals materials $MnBi_2Te_4$[116]. Another promising approach to overcoming these issues is to incorporate TIs in magnetic heterostructures, which allow for tuning of the magnetic properties independent of the dopants. Magnetic heterostructures open exchange gaps in TIs with reduced disorder compared to doped systems, as magnetic dopant and electric conduction are separated in space[117]. Integration of non-magnetic TIs with magnetically ordered materials can align spin moments and induce exchange splitting, similar to stacks with magnetic TIs. This promises a fruitful strategy for the systematic exploration of doping strategies in heterostructures.

*Realisation of spin states in unconventional semiconductors:* Quantum computing based on semiconducting quantum dots exploits the distinct spin states of confined electrons (See *Spin and topology-related phenomena in semiconductors*). Realising well-defined spin states and their control is an important technological challenge where the dissipation rate and coupling mechanisms of spins in exotic materials would largely differ from Si and GaAs. This underscores the appeal of graphene as a host for spin qubits[118], given that it offers long spin coherence times (see Box 1). Recent progress in gate-controlled quantum dots within bilayer graphene[119] brings the realisation of spin qubits within reach. Spin qubits in graphene facilitate exploration of the additional valley degree of freedom, presenting possibilities for novel spin-valley qubit implementation[120], beyond that of spin qubits in narrow and wide band gap semiconductors (see *Spin and topology-related phenomena in semiconductors*). 2D TMDs with large spin-orbit coupling and large band gaps provide effective spin modulation, while their inherent spin–valley locking can enable novel qubit designs and spintronic functionality[121].

*Topological quantum computation via Majorana zero modes (MZMs):* Despite recent claims of the experimental realisation of topological qubits based on MZMs[122], significant challenges remain in reliably creating and detecting these exotic states. A key issue is the development of unambiguous detection protocols for MZMs, as the currently employed TGP has been argued to produce inconsistent outcomes dependent on experimental parameters such as magnetic field range and junction transparency[123]. Material quality presents another significant hurdle, as the topological gap must be sufficiently large (around ~100 μeV) to suppress quasiparticle poisoning and decoherence effects and enable fault-tolerant quantum computation. Current experimental devices exhibit gap energies around at most ~50 μeV[124]. Addressing these challenges requires several key improvements. First, developing more robust detection protocols beyond the TGP would include establishing standardised data acquisition parameters, symmetric bias voltage ranges, and more transparent reporting of all regions passing the protocol criteria. Second, it would be worthwhile to investigate alternative experimental techniques that can directly probe the non-Abelian statistics of MZMs through braiding operations, which would provide definitive evidence beyond existing measurements. Finally, an unambiguous proof of MZMs as well as practical quantum operations will require further advances in material engineering to reduce disorder, increase superconducting gaps, and minimise quasiparticle poisoning rates. Independent verification across multiple research groups using standardised measurement protocols will be crucial to build confidence in reported advances.

**Perspective**

The most important roadblock on the path to QAHE-based quantum electronics is the materials, and intense research efforts are underway to improve the known and explore the unknown[117]. While it is important for $(Bi,Sb)_2Te_3$ TIs to follow the successful (but tedious) path of perfection that III-V materials have taken[125], improving Cr/V and Mn-based systems, *e.g.*, through heterostructure engineering[126] is promising. It is also equally important to experimentally explore the countless theoretically proposed topological electronic materials (*e.g*. $HoMn_6Sn_6$ and $Co_3Sn_2S_2$). Furthermore, QAHE has also been observed in twisted bilayer graphene at 1.6 K[127], making twisted materials a promising candidate for quantum metrology. The control of interlayer twist can be further exploited to tailor spin interactions in 2D magnets[128], enabling the perspective of 2D engineered topological magnetic structures like skyrmions (see *Magnetic Skyrmions*)[129]. Long-term anyon-based fault tolerant quantum computation might be realised by a 2D quantum system with anyonic excitations, such as superconductors or quantum spin liquids[6].

## *Outlook*

Materials innovation is emerging as a critical enabler for the scalable deployment of quantum technologies, particularly in regimes where system performance is constrained by coherence times, qubit connectivity, and cryogenic power budgets. Our analysis highlights specific scenarios where these limitations render progress in materials not merely advantageous, but fundamentally essential to overcoming current bottlenecks and enabling the next generation of quantum systems.

Advanced and hybrid material systems will also provide fertile ground for other spin-based quantum technologies, *e.g.* quantum magnonics where the quantum nature of collective spin excitations has been actively explored in hybrid quantum systems[130,131]. For example, superconducting qubits have been successfully coupled with magnon modes within a microwave cavity, realising the single magnon counting using qubit spectrum[132]. Very recently, using a similar magnon-qubit coupling scheme, nonclassical quantum superposition of the single-magnon and vacuum states has been demonstrated[133].

Alternatively, rare-earth ions in host crystals have emerged as leading candidates for solid-state quantum memories due to their unique electronic configurations, particularly the shielding of 4f electrons by filled outer 5s/5p orbitals, which weakens coupling to the host and enables long coherence times. Lanthanide-doped $Y_2SiO_5$ crystals such as $Eu:Y_2SiO_5$ or $Pr:Y_2SiO_5$ have demonstrated very long spin coherence times, making them ideal for long-lived qubit storage and quantum repeater applications. For example, in $Eu:Y_2SiO_5$[134], 10-hour coherence times have been reported. Recent advances include the demonstration of telecom-wavelength heralded entanglement between two spatially separated spin-wave quantum memories using rare-earth doped crystals, achieving temporal multiplexing across 15 modes[135] as well as notable progress in ensemble-based storage and on-demand retrieval. These developments position rare-earth ions as robust quantum memory platforms enabling scalable quantum networks. For more detailed reviews, the readers are recommended to check Refs. 136 and 137.

For topological and 2D systems, near-term priorities include increasing induced topological gaps and quasiparticle-poisoning times in superconductor-proximitised topological hybrids, establishing robust zero-field QSH platforms with superconducting proximity, and delivering reproducible 2D spin/valley qubits with controlled valley splitting.

Translating research into reliable products requires a supply of high quality, well-characterised materials. In this Perspective article, we have outlined exciting prospects for the various spin and topology material platforms. For example, impressive strides are being made in the deterministic creation of spin qubits in semiconductors and minimising noise, while superconductors hold promise for useful emergent phenomena at semiconductor and magnetic interfaces. Topology presents intriguing possibilities in terms of both skyrmion qubits and MZMs that may be adopted for QT in a longer-term future. Production and characterisation of materials for QT requires capability in terms of advanced techniques and infrastructure, *e.g.* advanced nanofabrication, control of noise and readout, advanced readout schemes for various types of spin- and topology-based qubits, including those based on skyrmions and MZMs. Such advanced production and characterisation capabilities as well as relevant challenges (as outlined in this article) are often common for different types of materials, making sense to combine the effort, sharing the knowledge and finding a common solution.

We have a firm belief that the advance in quantum materials is a cornerstone in the success of QT applications, and the relevant R&D work supported by the national/international infrastructure, regulations and policies, standards and benchmarking will fully develop the industrial potential of such applications and bring wide economic and societal benefit.

Realising the technological benefits of quantum science will require a truly interdisciplinary effort of material scientists, quantum physicists and engineers. By providing a visionary (but still realistic) approach, we believe that this Perspective article helps to bring together and strengthen the quantum community.

### *Supplementary Material*

The Supplementary Material contents a detailed information about i) Semiconducting spin qubits and their individual figures of merit (Table S1) and ii) Comparison of critical FOMs for the development of skyrmion-based quantum technologies, also including the corresponding values for meron- and domain wall-based approaches (Table S2).

### *Acknowledgement*

This roadmap is conceived and delivered by the EPSRC Materials for Quantum Network (M4QN), which acts as a bridge between the UK and international material science and quantum communities to bring disruptive new insight to QTs. The authors acknowledge the support of the UK Government Department for Science, Innovation and Technology through the UK National Quantum Technologies Programme. The authors are supported by the Engineering and Physical Sciences Research Council (EPSRC) under the grants EP/V028189/1 and EP/X015661/1; UKRI Future Leaders Fellowship MR/Y017331/1; Leverhulme Trust via fellowship (RF-2024-317); the Horizon Europe Project “2D Heterostructure Non-volatile Spin Memory Technology" (2DSPIN-TECH) supported by UKRI grant number [10101734] and the Royal Commission of 1851 Research Fellowship. The authors are grateful to Tobias Lindstrom and Alexander Tzalenchuk (NPL) for useful comments and discussions.

### *Author Contribution*

OK conceived the idea and led the delivery of the roadmap. All authors have contributed to the writing of the manuscript.